\documentclass[useAMS,usenatbib,fleqn]{mnras}
\usepackage{graphicx,subfig,lscape}
\usepackage[usenames,dvipsnames]{xcolor}
\usepackage{times}

\newcommand{\kms}{km\,s$^{-1}$}

\newcommand{\firxps}[1]{\resizebox{0.8\hsize}{!}{\rotatebox{-90}{\includegraphics{#1}}}}
\newcommand{\fifps}[2]{\centering\resizebox{#1}{!}{\includegraphics{#2}}}
\newcommand{\firrps}[2]{\resizebox{#1}{!}{\rotatebox{270}{\includegraphics{#2}}}}

\title[Testing time-dependent atomic diffusion theory]{Putting atomic diffusion theory of magnetic ApBp stars to the test: 
evaluation of the predictions of time-dependent diffusion models}

\author[O. Kochukhov and T. A. Ryabchikova]
{O. Kochukhov$^{1}$ and T. A. Ryabchikova$^{2}$\\
$^{1}$Department of Astronomy and Space Physics, Uppsala University, 751 20 Uppsala, Sweden\\
$^{2}$Institute of Astronomy of the Russian Academy of Sciences, 48 Pyatnitskaya St., 119017 Moscow, Russia
}  

\begin{document}

\date{Accepted 2017 November 7. Received 2017 November 5; in original form 2017 September 29}

\pagerange{\pageref{firstpage}--\pageref{lastpage}} \pubyear{2017}

\maketitle

\label{firstpage}

\begin{abstract}
A series of recent theoretical atomic diffusion studies has address the challenging problem of predicting inhomogeneous vertical and horizontal chemical element distributions in the atmospheres of magnetic ApBp stars. Here we critically assess the most sophisticated of such diffusion models -- based on a time-dependent treatment of the atomic diffusion in a magnetised stellar atmosphere -- by direct comparison with observations as well by testing the widely used surface mapping tools with the spectral line profiles predicted by this theory. We show that the mean abundances of Fe and Cr are grossly underestimated by the time-dependent theoretical diffusion model, with discrepancies reaching a factor of 1000 for Cr. We also demonstrate that Doppler imaging inversion codes, based either on modelling of individual metal lines or line-averaged profiles simulated according to theoretical three-dimensional abundance distribution, are able to reconstruct correct horizontal chemical spot maps despite ignoring the vertical abundance variation. These numerical experiments justify a direct comparison of the empirical two-dimensional Doppler maps with theoretical diffusion calculations. This comparison is generally unfavourable for the current diffusion theory, as very few chemical elements are observed to form overabundance rings in the horizontal field regions as predicted by the theory and there are numerous examples of element accumulations in the vicinity of radial field zones, which cannot be explained by diffusion calculations.
\end{abstract}


\section{Introduction}
\label{intro}

The upper main-sequence chemically peculiar stars are distinguished by a bewildering diversity of surface chemical anomalies. Several groups of these stars exhibit conspicuous overabundances of iron-peak and heavy elements in comparison to the abundance patterns of the Sun and cool main-sequence stars. Many of these chemically peculiar stars possess strong, stable, globally-organised magnetic fields. These so-called magnetic ApBp stars exhibit the most extreme abundance anomalies, often accompanied by the horizontal \citep[e.g.][and references therein]{kochukhov:2017} and vertical \citep[e.g.][and references therein]{ryabchikova:2014} chemical abundance inhomogeneities.

It is generally thought that non-solar surface abundances in the upper main-sequence stars arise due to selective levitation and sinking of chemical elements under the competing influence of radiation pressure and gravity. This atomic diffusion hypothesis \citep*{michaud:1970,michaud:2015} has been reasonably successful in explaining the (horizontally homogeneous) build-up of metals in the radiative interiors of old Sun-like stars \citep*{richard:2005} and metallic-line A-type stars \citep{vick:2010}. In comparison, atmospheric diffusion calculations required to explain large element overabundances in magnetic ApBp stars and, especially, their highly non-uniform surface abundance topologies, are considerably more uncertain and computationally challenging. This is due to the necessity of detailed treatment of the radiative transport in a magnetised stellar atmosphere and uncertainties concerning the potential impact of several poorly constrained effects, such as weak stellar winds, on the diffusion calculations.

The first self-consistent equilibrium (based on the assumption of zero net particle flux) atmospheric diffusion models were presented by \citet*{hui-bon-hoa:2000} for blue horizontal-branch stars. Later, these models were improved and applied to chemically peculiar stars \citep{leblanc:2004,leblanc:2009}. These calculations, predicting large mean element overabundances as well as substantial vertical chemical gradients in the line-forming atmospheric regions, compare favourably with observations \citep{ryabchikova:2008a,ryabchikova:2011a,ryabchikova:2011}. On the other hand, since these models did not fully incorporate a magnetic field, they could not be used to predict the lateral variation of chemical stratification across the stellar surface according to the local field strength and inclination. Nevertheless, \citet{leblanc:2009} showed that the effect of reduced ion mobility in a transverse magnetic field should lead to an additional element accumulation in the upper part of the stellar atmosphere, in the vicinity of surface regions characterised by highly inclined magnetic field lines.

\citet{alecian:2010} and \citet{alecian:2015} presented an independent set of equilibrium diffusion calculations, incorporating detailed treatment of a magnetic field. The bi-dimensional element distributions published by \citet{alecian:2010} confirmed the hypothesis by \citet{leblanc:2009} that the main effect of an inclined magnetic field is to enhance element concentration in the upper part of the stellar atmosphere. Surprisingly, all chemical elements were predicted to share essentially the same behaviour with respect to the magnetic field geometry, leading to ubiquitous element overabundance rings at the magnetic equators of ApBp stars with dipole-like fields. As pointed out by several observational studies (e.g. \citealt*{silvester:2015}; \citealt{kochukhov:2017a}), these predictions are at clear odds with the diversity of surface chemical spot topologies recovered by Doppler imaging (DI) studies of ApBp stars and are incompatible with some well-established examples of overabundance spots coinciding with the radial field regions at the magnetic poles (\citealt{kochukhov:2004e}; \citealt*{silvester:2014a}).

A new generation of magnetic atomic diffusion calculations was recently developed by \citet{alecian:2011}, \citet{stift:2016}, and \citet{alecian:2017}. Unlike the previous equilibrium diffusion models, these new calculations rely on the assumption of a constant particle flux throughout the stellar atmosphere. The steady-state chemical stratification solutions inferred by these models are significantly different from the corresponding results of the equilibrium diffusion calculations, with important consequences for the resulting three-dimensional element abundance distributions and, presumably, for interpretation of observations. Moreover, \citet[][hereafter SA16]{stift:2016} suggested that the complex 3D nature of their theoretical element distributions undermines the validity of widely used 2D abundance mapping tools (Doppler and magnetic Doppler imaging), which assume vertically constant element abundances. 

However, neither this particular conjecture nor the general predictions of the time-dependent atmospheric diffusion theory have been ever tested by comparison with observations or by applying common spectral analysis procedures to line profiles simulated according to this theory. In this paper we present both types of analysis. We start by explaining the reasoning behind choosing the chemical stratification profiles from the paper by SA16 among several available diffusion calculations (Sect.~\ref{models}). We then describe our own model atmosphere and line profile calculations using theoretical chemical stratification data (Sect.~\ref{calcs}). We proceed with the analysis of theoretical predictions by first establishing the mean element abundances from theoretical models consistently with the abundance measurements of real ApBp stars (Sect.~\ref{amean}) and then exploring the response of 2D surface mapping codes to the line profiles calculated for a theoretical 3D abundance structure (Sects.~\ref{indi} and \ref{lsd}). The paper ends with conclusions and discussion in Sect.~\ref{disc}, where we summarise the results of our analysis and consider the predictions of theoretical models in the context of observations of real magnetic ApBp stars.

\section{Theoretical diffusion models}
\label{models}

\begin{figure*}
\centering
\firrps{15.2cm}{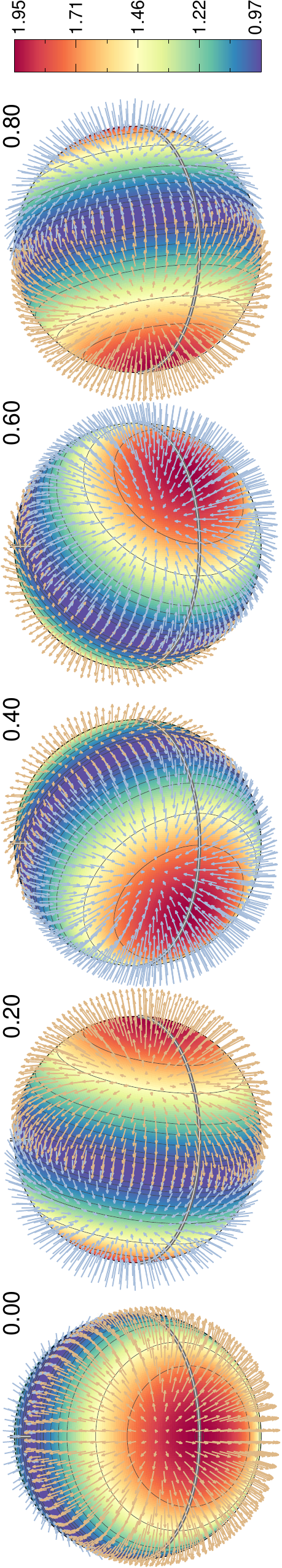}\vspace*{0.2cm}
\firrps{15cm}{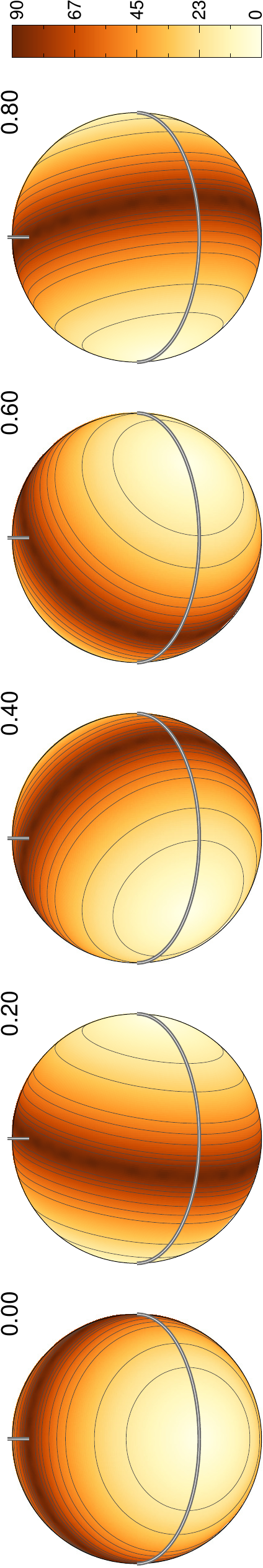}
\caption{Dipolar magnetic field topology adopted in our calculations. The upper row of spherical plots shows the field orientation (outward and inward pointing vectors) plotted over the field modulus distribution (colour map). The field strength scale, in kG, is indicated by the side bar. The lower set of spherical plots shows the corresponding field inclination with respect to the local surface normal. In this case, the side bar gives the absolute value of the field inclination angle in degrees. The solid contour lines correspond to the 9 field inclinations considered by SA16. The star is shown at five rotational phases, as indicated above each panel. The thick double line and the short bar indicate the stellar rotational equator and the pole, respectively.}
\label{fld}
\end{figure*}

\begin{figure*}
\centering
\firrps{15cm}{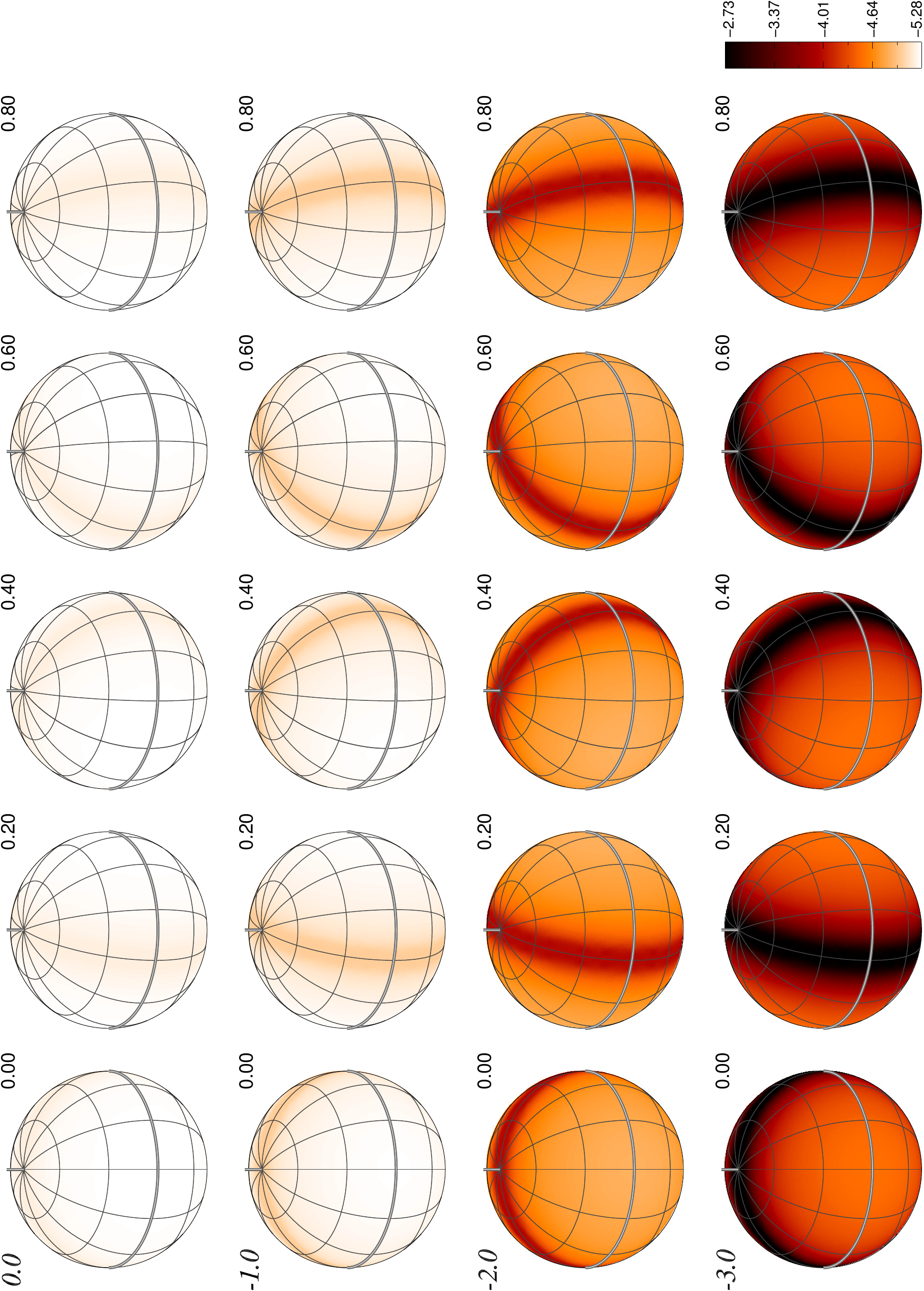}
\caption{Surface distribution of Fe abundance at the optical depths $\log \tau_{5000}=0.0$, $-1.0$, $-2.0$, and $-3.0$ (indicated with labels at left) for the time-dependent diffusion model of SA16. The abundance scale, in the $\log N_{\rm el}/N_{\rm tot}$ units, is indicated by the colour bar at right.}
\label{strat_fe}
\end{figure*}

\begin{figure*}
\centering
\firrps{15cm}{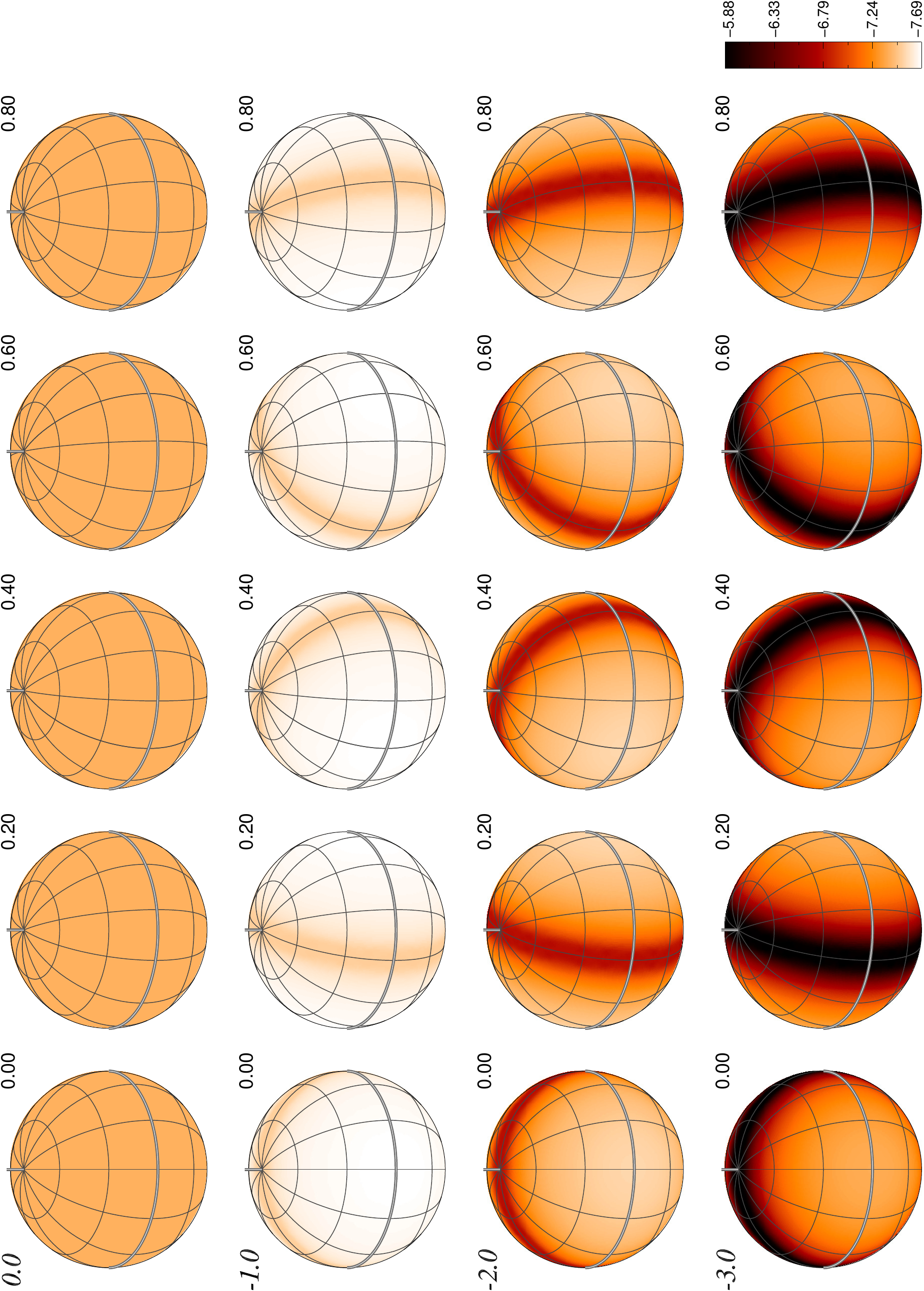}
\caption{Same as Fig.~\ref{strat_fe} for Cr.}
\label{strat_cr}
\end{figure*}

Not every theoretical atomic diffusion calculation which has appeared in the recent literature lends itself to a straightforward quantitative analysis and comparison with observations. The comprehensive presentation of the bi-dimensional stratification profiles by \citet{alecian:2010} provided data for 16 chemical elements and 4 values of effective temperature. However, these authors did not include information on the physical height in the stellar atmosphere in their plots, limiting their usefulness for an independent quantitative analysis. They also presented calculations only for a 20~kG dipolar magnetic field, which is well in excess of the 2.5~kG mean dipolar field strength of the nearby ApBp stars \citep{power:2008} and the 2.2~kG mean field of the Ap stars typically targeted by Doppler imaging studies \citep{kochukhov:2017}. In addition, the bi-dimensional abundance maps presented by \citet{alecian:2010} were based on only three individual stratification profiles at different field inclinations. Subsequent higher angular resolution calculations by \citet{alecian:2012} revealed a significantly more structured surface abundance distribution, corresponding, in the case of Fe, to an equatorial overabundance ring of only $\pm4\degr$ width. 

Considering the newer time-dependent diffusion models, \citet{alecian:2017} presented three-dimensional abundance distributions of Cr and Fe in a putative offset dipolar magnetic field geometry corresponding to that of the extraordinarily magnetic, cool Ap star HD\,154708 \citep{hubrig:2005}. This object can hardly be considered representative of early-A magnetic chemically peculiar stars, which currently provide the bulk of observational constraints for the atomic diffusion theory. Furthermore, its low projected rotational velocity and an unfavourable inclination preclude deriving detailed magnetic and chemical spot maps which could be usefully compared to the model predictions.

This leaves us with the study by SA16, who presented the predicted vertical dependence of the Cr, Fe, and V abundance across the surface of a typical Ap star ($T_{\rm eff}=10000$~K, $\log g=4.0$) with a 1.95~kG centred dipolar magnetic field. The vertical abundance profiles were calculated for 9 different magnetic co-latitudes, enabling fine sampling of the stellar surface. By virtue of choosing representative stellar parameters and tabulating the results at a sufficiently dense angular grid, these calculations are well suited for comparison with observations of typical Ap stars in this temperature range and for testing empirical abundance mapping tools. We exploit those theoretical (3D) distributions to compute synthetic spectral timeseries, which we in turn employ to reconstruct the implied surface distribution as viewed through the filter of (2D) DI.

However, no stellar rotation or other surface effects were considered in the calculations by SA16. Consequently, the orientation of the dipolar field axis and associated chemical abundance maps with respect to the stellar rotational axis is arbitrary. In this study we assumed a large obliquity angle ($\beta=90\degr$) and a moderately large inclination angle ($i=60\degr$), resulting in a reversing longitudinal magnetic field and significant line profile variation. The adopted magnetic field geometry is illustrated in Fig.~\ref{fld}. Figures~\ref{strat_fe} and \ref{strat_cr} present the horizontal abundance maps of Fe and Cr at four representative optical depths. These maps were obtained with the help of linear interpolation within the grid of 9 vertical abundance profiles provided for each element by SA16 in their Fig.~5. It is evident that, despite being based on a significantly different physical foundation and different numerical treatment compared to the equilibrium approach to the atmospheric diffusion, the time-dependent diffusion model predicts a qualitatively similar accumulation of chemical elements, which coincides with the magnetic equator region and grows with height in the stellar atmosphere.

\section{Model atmosphere and line profile calculations}
\label{calcs}

Using the set of Fe and Cr stratification profiles reported by SA16, we calculated a grid of 9 {\sc LLmodels} \citep{shulyak:2004} atmospheres for $T_{\rm eff}=10000$~K and $\log g=4.0$. These calculations were iterated several times to achieve full consistency between the stratification specified on the $\log\tau_{5000}$ scale and the model $T(\tau)$ relation. Abundances of other chemical elements were assumed to be solar and their vertical distributions were taken to be homogeneous. As shown by \citet{khan:2007}, Fe and Cr are by far the most important sources of line opacity in ApBp-star atmospheres. SA16 also showed that there is little back-reaction of the stratification of other elements on that of Fe, implying that individual abundances and stratifications of other elements are relatively unimportant for the model atmosphere structure. In addition to chemically stratified models, we also calculated a reference solar abundance model for $T_{\rm eff}=10000$~K and $\log g=4.0$.

In the next step, we employed the {\sc Invers13} code \citep{kochukhov:2013} to simulate a set of Fe~{\sc ii} spectral line profiles for the theoretical 3D abundance distribution of Fe. Originally developed for a self-consistent magnetic field and temperature mapping, {\sc Invers13} can be used in the forward mode to synthesise the Stokes parameter spectra for an arbitrary surface map relating a certain scalar parameter to a grid of local model atmospheres. In this case, the model grid incorporates different Cr and Fe abundance stratifications as a function of the magnetic co-latitude. Using {\sc Invers13}, we calculated profiles of the intermediate strength Fe~{\sc ii} lines 4273.32, 4520.22, 4555.89, 4666.75~\AA, used by \citet{silvester:2014a} in their magnetic DI study of $\alpha^2$~CVn. The Zeeman effect due to the (vertically uniform) 1.95~kG dipolar magnetic field was included in detail, with the individual Zeeman splitting patterns and other parameters of the studied transitions extracted from the VALD3 data base \citep{ryabchikova:2015}. The Fe line profiles were calculated for 20 equidistant rotational phases, assuming the projected rotational velocity of $v\sin i=25$~\kms\ and the resolving power of $\lambda/\Delta\lambda=120000$, corresponding to the HARPS polarimeter \citep{piskunov:2011}. Random noise with $\sigma=2\times10^{-3}$ (signal-to-noise ratio 500) was added to the data. These spectra will be used in Sect.~\ref{indi} to test the individual line abundance mapping procedure.

Another set of non-magnetic local Stokes $I$ spectra was calculated for each of the 9 chemically stratified model atmospheres with the {\sc Synmast} \citep*{kochukhov:2010a} code, this time considering all metal lines deeper than 1\% of the continuum in the 4000--7000~\AA\ wavelength interval. The local intensity profiles were tabulated for a grid of 20 limb angles. Using this spectral library, we performed the surface integration according to the field geometry in Fig.~\ref{fld} with the help of custom IDL routines, obtaining normalised flux spectra for 20 equidistant rotational phases and two values of $v\sin i$, 0 and 25~\kms. The instrumental broadening corresponding to the resolving power of 120000 was applied to both sets of calculations. The resulting zero $v\sin i$ spectra will be used in Sect.~\ref{amean} for the determination of mean Fe and Cr abundances. The $v\sin i=25$~\kms\ spectra were further modified by adding random noise with $\sigma=1\times10^{-2}$ (signal-to-noise ratio 100). These spectra will be used for the least-squares deconvolution (LSD) analysis and inversion in Sect.~\ref{lsd}.

\section{Analysis of theoretical 3D abundance maps}
\label{analysis}

\subsection{Mean abundances}
\label{amean}

\begin{figure*}
\centering
\firrps{15cm}{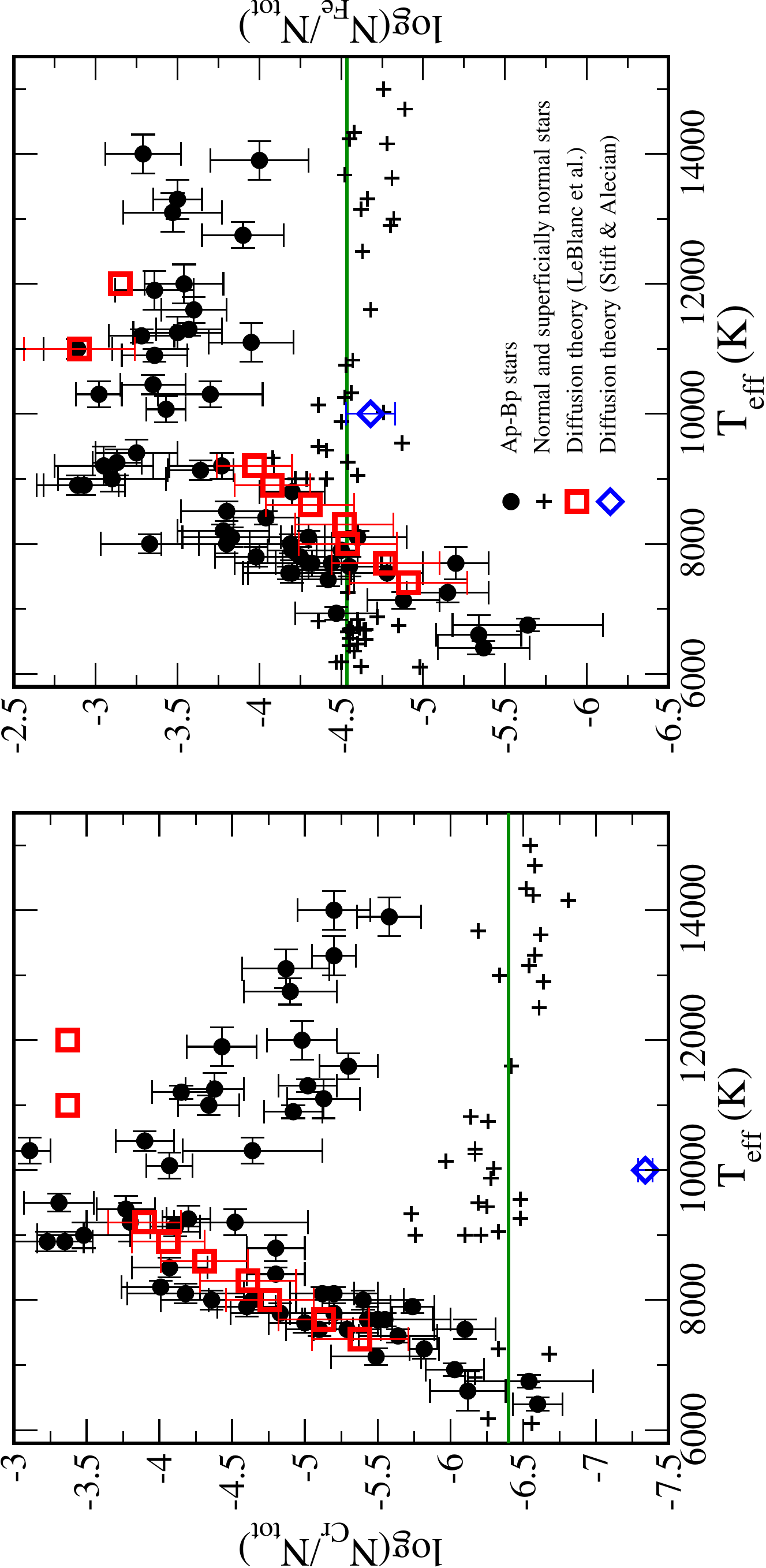}
\caption{Mean abundances of Cr (left panel) and Fe (right panel) for ApBp (circles) and normal (plusses) stars of different $T_{\rm eff}$ \citep{ryabchikova:2005b}. Diffusion  predictions are shown with squares for the models by \citet{leblanc:2004} and \citet{leblanc:2009} and with rhombs for the SA16 model analysed in this paper. All abundances are given in the $\log N_{\rm el}/N_{\rm tot}$ scale. The horizontal solid lines indicate solar abundances.}
\label{abund}
\end{figure*}

\begin{figure*}
\centering
\firxps{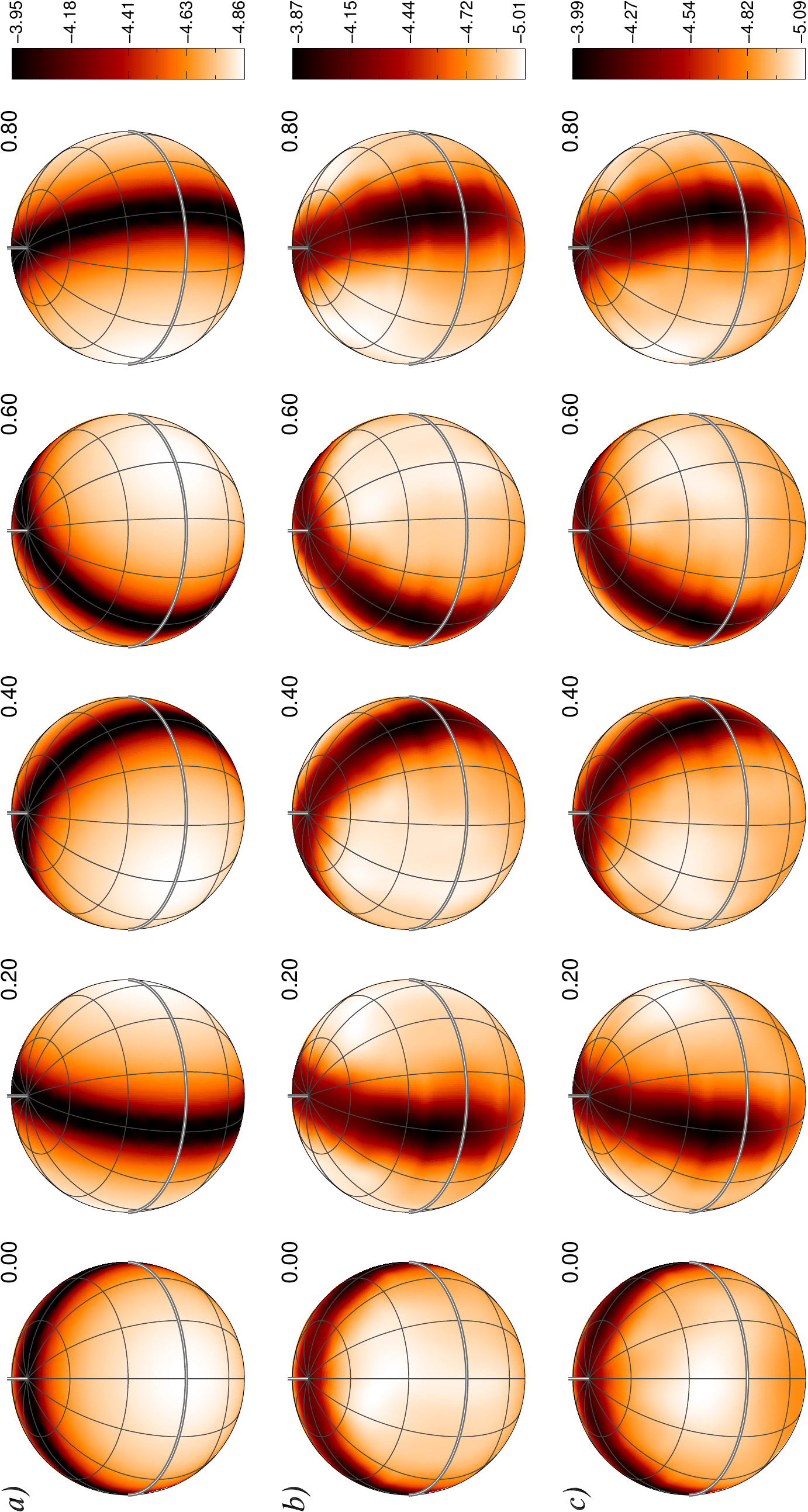}
\caption{Horizontal distribution of Fe abundance at the representative line formation optical depth ($\log \tau_{5000}=-2$) from the SA16 theoretical 3D diffusion model (a) compared to the 2D Doppler imaging reconstruction using individual Fe~{\sc ii} lines (b) and the iron LSD profiles (c). The abundance scale, in the $\log N_{\rm el}/N_{\rm tot}$ units, is indicated by the colour bar at right.}
\label{abn3d-2d}
\end{figure*}

We used the zero $v\sin i$ flux profiles for rotational phases 0.0 (minimum strength of Fe and Cr lines) and 0.25 (maximum Fe and Cr line strength), as well as the phase-averaged spectrum for the classical abundance determination with the equivalent width method. Initially, we aimed to perform this analysis based on the same set of neutral and singly ionised lines as was employed by \citet{ryabchikova:2014a} for the analysis of the non-magnetic equilibrium diffusion calculations by \citet{leblanc:2009}. However, it turned out that, with the iron distribution from the study by SA16, only 9 lines from that list (Fe~{\sc i} 5383.37~\AA\ and Fe~{\sc ii} 5410.91, 4923.92, 5132.66, 5169.03, 5197.58, 5291.66, 5325.55, 6432.68~\AA) were strong enough for a meaningful abundance estimate. The mean Fe abundance inferred from these spectral features using the reference solar-abundance model atmosphere is $\log N_{\rm Fe}/N_{\rm tot}=-4.77\pm0.12$ for phase 0.0, $-4.61\pm0.17$ for phase 0.25, and $-4.68\pm0.15$ for the phase-averaged spectrum. The studied lines cover a wide range of excitation potentials, from 2.8 to 10.5~eV, and therefore yield slightly different abundances depending on their formation depth. This is reflected in the quoted standard deviations. On the other hand, the range of mean abundance variation with rotational phase is only 0.09~dex.

Determination of the mean Cr abundance was based on equivalent widths of 8 Cr~{\sc ii} lines (4261.91, 4558.65, 4554.99, 4588.20, 4592.05, 4618.80, 4634.07, 4824.13~\AA). These are some of the strongest Cr features in the spectra of ApBp stars and are often unsuitable for an abundance analysis due to their saturation. However, for the 3D Cr abundance distribution published by SA16 the equivalent widths of these lines do not exceed 25~m\AA. Using these Cr~{\sc ii} transitions, we derived $\log N_{\rm Cr}/N_{\rm tot}=-7.39\pm0.02$ for phase 0.0, $-7.30\pm0.01$ for phase 0.25, and $-7.34\pm0.01$ for the phase-averaged spectrum. All considered Cr~{\sc ii} lines have excitation potentials in a narrow range from 3.9 to 4.1~eV and therefore similar line formation depth, which explains the small line-to-line scatter. 

In Fig.~\ref{abund} we compare the phase-averaged Fe and Cr abundances inferred by our analysis of the SA16 model with observed Fe and Cr abundances for the sample of normal and chemically peculiar ApBp stars compiled by \citet{ryabchikova:2005b}, and with the mean abundances corresponding to the non-magnetic diffusion models by \citet{leblanc:2004} and \citet{leblanc:2009} (see \citet{ryabchikova:2008a} and \citet{ryabchikova:2014a} for details). It is evident that while the latter theoretical models achieve a reasonable agreement with observations in a wide $T_{\rm eff}$ range, the time-dependent diffusion calculations by SA16 underestimate the apparent mean Fe abundance at $T_{\rm eff}=10000$~K by about 1.5 dex and underpredict the mean Cr abundance by as much as 3~dex.

\begin{figure}
\centering
\fifps{6.9cm}{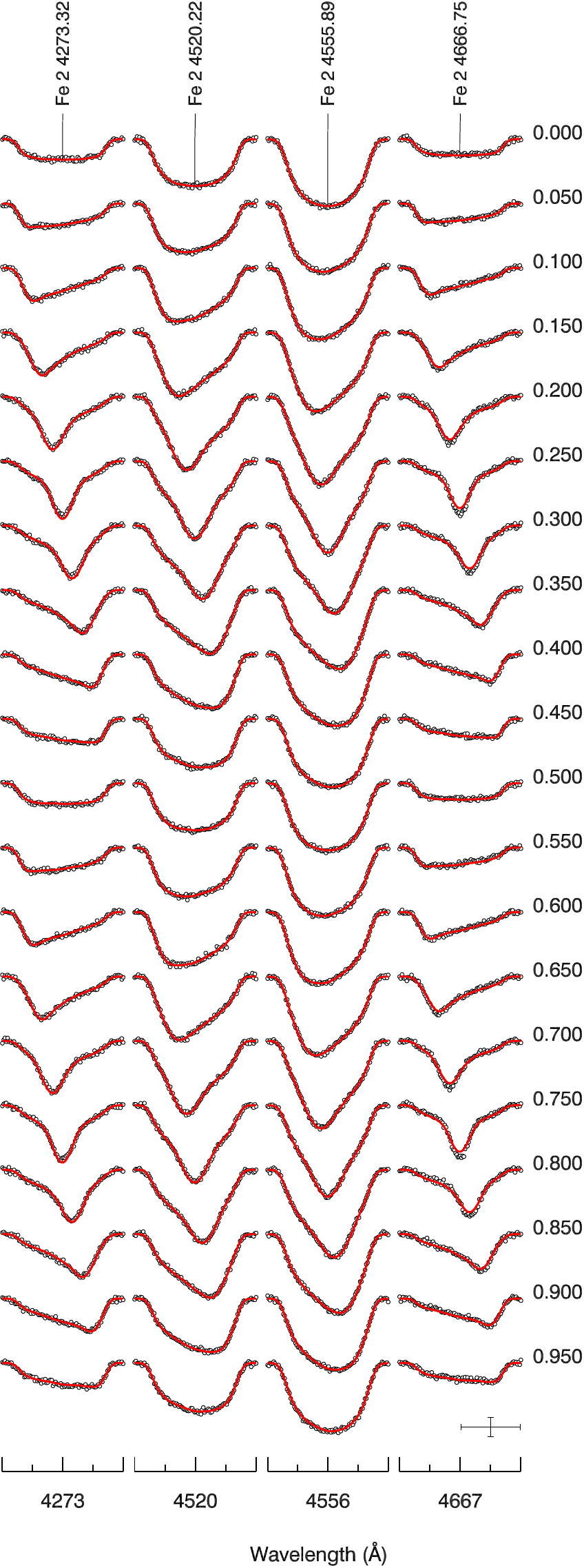}
\caption{Comparison of the Stokes $I$ profiles of individual Fe~{\sc ii} lines simulated according to the 3D atomic diffusion model (symbols) and the fit achieved by the {\sc Invers10} code (solid line) with the 2D Fe abundance map illustrated in Fig.~\ref{abn3d-2d}b. Rotational phases are indicated on the right side of the plot. The spectra are offset vertically. The error bars in the lower right corner indicate the horizontal (0.5~\AA) and vertical (3\%) scales.}
\label{prf_i10}
\end{figure}

\begin{figure}
\centering
\fifps{6cm}{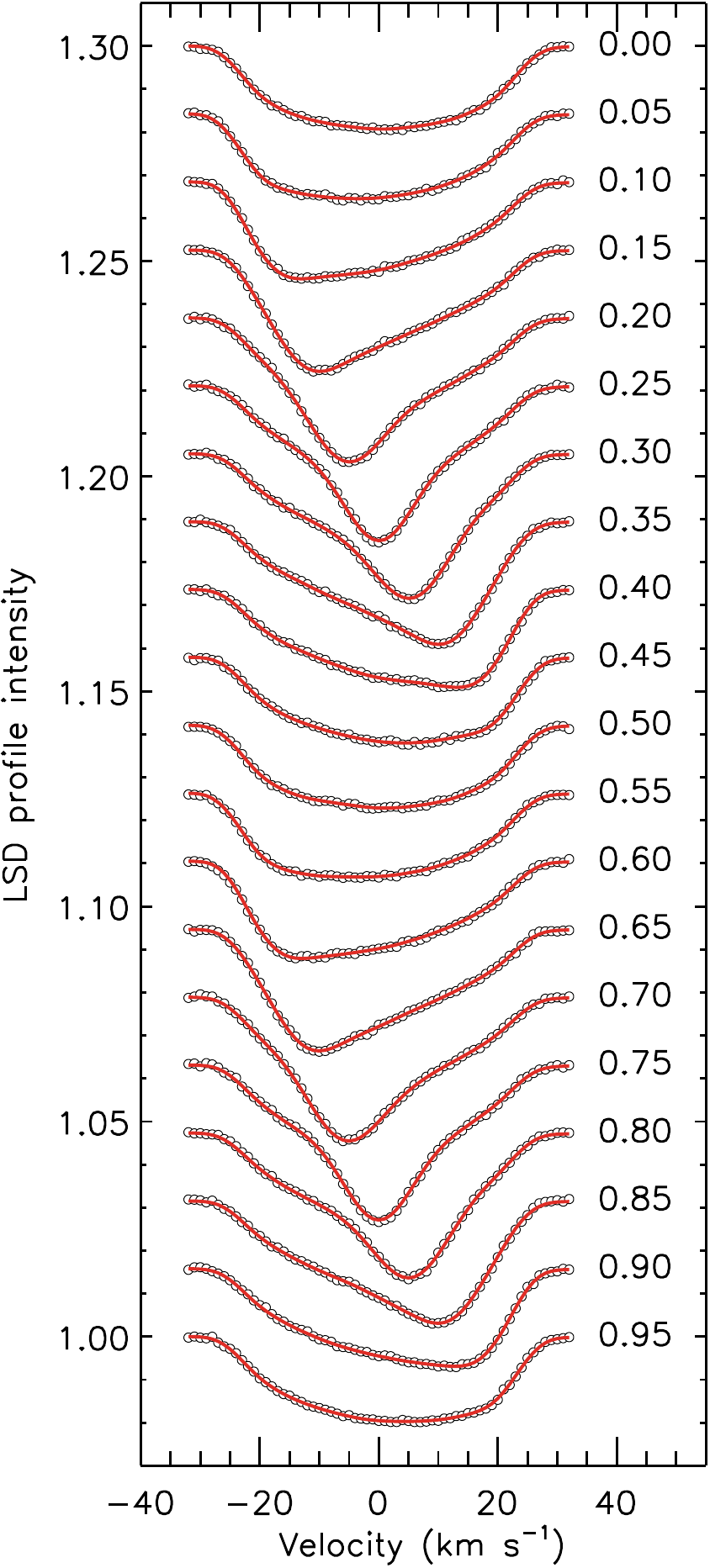}
\caption{Same as Fig.~\ref{prf_i10} for the comparison of the Stokes $I$ Fe LSD profiles simulated according to the 3D atomic diffusion model (symbols) and the fit achieved by the {\sc InversLSD} code (solid line) with the 2D Fe abundance map illustrated in Fig.~\ref{abn3d-2d}c.}
\label{prf_lsd}
\end{figure}

\subsection{Doppler imaging with individual lines}
\label{indi}

In this test we used the Stokes $I$ profiles of the four Fe~{\sc ii} lines computed as described in Sect.~\ref{calcs} for mapping the horizontal Fe abundance distribution with the {\sc Invers10} code \citep{piskunov:2002a,kochukhov:2002c}. The fixed dipolar magnetic field was taken into account. This inversion was carried out under the usual assumption of a vertically homogeneous Fe abundance distribution and relied upon a single, solar-abundance model atmosphere. Following the practice of many recent multi-line abundance DI studies, we allowed the code to adjust oscillator strengths of three out of four lines. In this case, such an adjustment partly compensates for the chemical stratification effects (which are not included in our model). In applications to real stars, this oscillator strength correction also alleviates inevitable errors in atomic data and in treatment of unresolved blends.

The DI inversion was constrained with the Tikhonov regularisation function. The corresponding regularisation parameter was chosen according to the procedure described by \citet{kochukhov:2017}. The resulting fit to 20 phases of simulated observations is presented in Fig.~\ref{prf_i10}. {\sc Invers10} succeeds in reproducing the spectral line profiles corresponding to the theoretical 3D Fe abundance distribution with a 2D map. The final standard deviation of the fit (0.22\%) is only marginally larger than the nominal random noise (0.2\%) injected in the data. The $\log gf$ corrections do not exceed 0.06~dex. The 2D Fe map obtained by {\sc Invers10} is shown in Fig.~\ref{abn3d-2d}b. This distribution turns out to be very similar to the cross-section of the input 3D element distribution at $\log\tau_{\rm 5000}=-2$ (Fig.~\ref{abn3d-2d}a), which corresponds to the typical formation depths of intermediate-strength Fe lines. The morphology of the principal surface features and the range of Fe abundance variation across the stellar surface agree very well in these two maps.

\subsection{Doppler imaging with least-squares deconvolved profiles}
\label{lsd}

In the final test we assessed the reconstruction of the Fe horizontal abundance map from the Stokes $I$ LSD profiles. This experiment is motivated by the common application of such mean profiles for simultaneous mapping of the magnetic and abundance structures of ApBp stars \citep{folsom:2008,kochukhov:2014,kochukhov:2017a,oksala:2017}.

The input data for the inversion was calculated by applying the {\sc iLSD} code \citep{kochukhov:2010a} to 20 rotational phases of $v\sin i=25$~\kms\ spectra calculated for the wide wavelength interval as described in Sect.~\ref{calcs}. The LSD line mask was comprised of  566 Fe~{\sc I} and {\sc II} lines deeper than 5\% of the continuum, as well as 249 blending features of other chemical elements treated with a separate mean profile. The LSD procedure increased the signal-to-noise ratio from 100 per 0.8~\kms\ velocity bin in the input spectra to about 2300 per 1.0~\kms\ bin in the mean profiles. The resulting Fe LSD profiles were interpreted with the {\sc InversLSD} surface mapping code, based on the methodology described by \citet{kochukhov:2014}. This inversion code models the intensity or/and polarisation LSD spectra using a pre-computed grid of local mean profiles. This grid was calculated by applying LSD to the local spectra synthesised for a range of vertically homogeneous Fe abundances, using a single, reference solar abundance model atmosphere.

The final model fit to the simulated LSD profiles is shown in Fig.~\ref{prf_lsd}. The corresponding 2D Fe distribution inferred by {\sc InversLSD} is illustrated in Fig.~\ref{abn3d-2d}c. Once again, we find a very good concordance between this LSD-based Fe map, the results of individual line mapping discussed in Sect.~\ref{indi}, and a horizontal cross-section of the theoretical 3D iron abundance distribution.

\section{Conclusions and discussion}
\label{disc}

In this paper we have carried out a detailed assessment of the predictions of recent theoretical time-dependent atomic diffusion calculations by \citet{stift:2016}. Their model aims to realistically describe the overall accumulation of metals, as well as the vertical and horizontal chemical inhomogeneities, in the atmospheres of magnetic ApBp stars. Deriving mean Fe and Cr abundances from the spectra generated according to the three-dimensional theoretical distributions of these elements, we revealed a major discrepancy between theoretically predicted and observationally inferred mean element concentrations. Contrary to the well-established $\sim$\,1~dex Fe and $\sim$\,2~dex Cr average overabundances typical of Ap stars with $T_{\rm eff}$ close to 10000~K, the model by SA16 predicts a near-solar mean abundance of Fe and $\sim$\,1~dex \textit{underabundance} of Cr. 

\begin{figure}
\centering
\fifps{8cm}{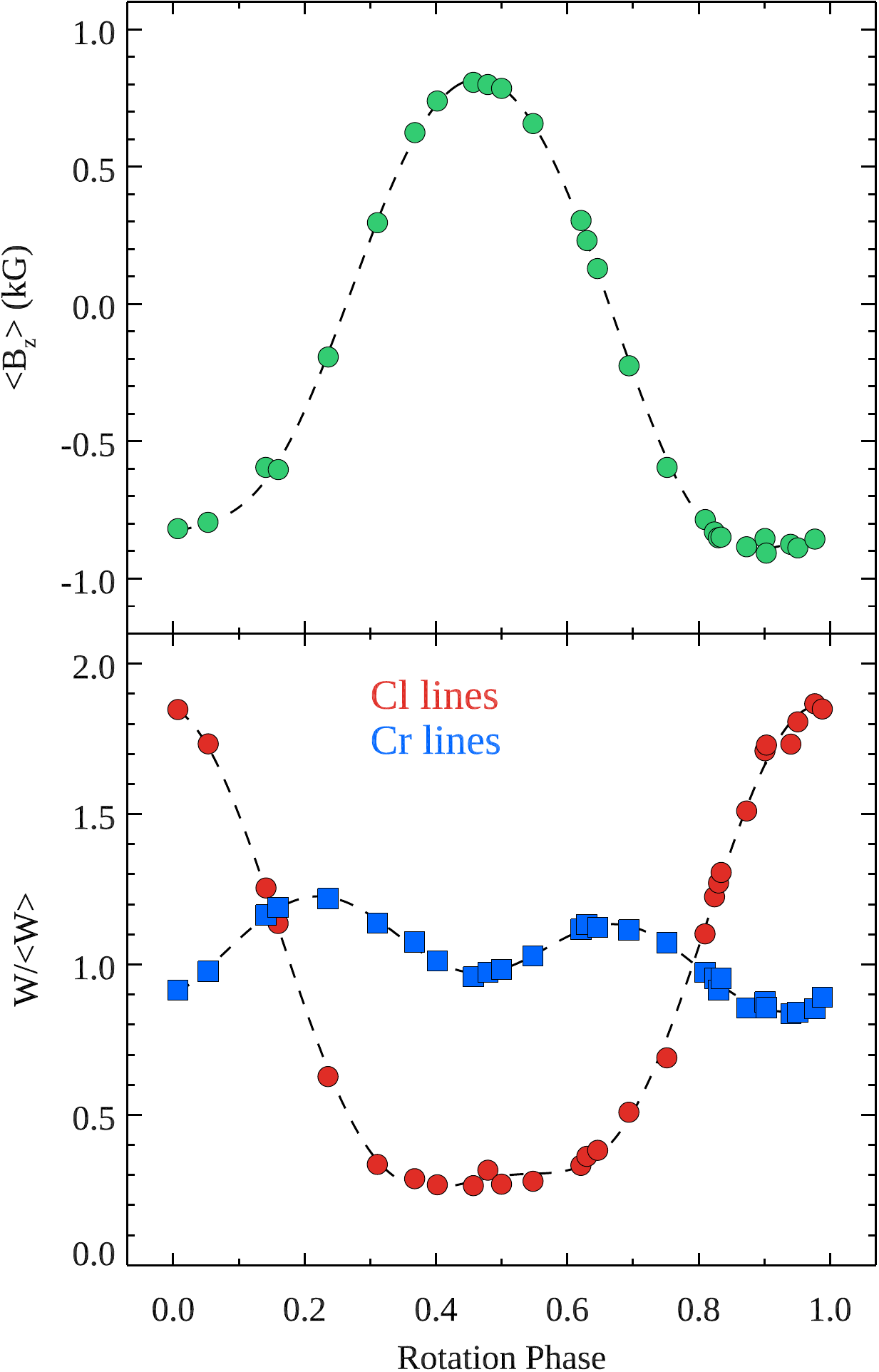}
\caption{Variation of the mean longitudinal magnetic field (upper panel) and relative equivalent widths (lower panel) of Cl (circles) and Cr (squares) lines in the spectrum of the prototypical Ap star $\alpha^2$~CVn.}
\label{ewbz}
\end{figure}

To the best of our knowledge, the inability of time-dependent diffusion calculations to predict correct mean Fe-peak abundances has not been previously discussed in the literature. Considering the presentation by \citet{alecian:2010} of the bi-dimensional element distributions, one can conclude that the problem of underpredicting abundances of certain element (e.g. Si) was also present to some extent in their earlier equilibrium models. On the other hand, Cr and Fe were estimated to be overabundant at $T_{\rm eff}=10000$~K \citep[see Figures~14 and 15 of][]{alecian:2010}, in agreement with independent non-magnetic atomic diffusion modelling by \citet{leblanc:2009}. The latter calculations have been extensively verified by comparison with observations and were shown to yield correct mean Fe-peak element abundances, to reproduce their trends with $T_{\rm eff}$, and to provide a reasonable match to the observed vertical element distributions \citep{ryabchikova:2014a,ryabchikova:2014}. In this context, the prediction of completely unrealistic mean abundances by the newest time-dependent version of the atomic diffusion theory is a step backwards. This problem needs to be addressed before these theoretical models can be considered to be a competitive contribution to our understanding of these phenomena.

In addition to examining mean Fe and Cr abundances, we performed a series of numerical simulations to study the robustness of the standard 2D Doppler imaging analysis applied to the line profiles computed according to 3D theoretical element distributions. These tests demonstrated that DI mapping, whether based on modelling of individual spectral lines or on interpretation of LSD profiles, yields sensible two-dimensional element distributions with the surface structure details and abundance ranges closely matching the input 3D model at the typical line-forming atmospheric depth. These results provide a clear demonstration that vertical abundance gradients do not confuse DI codes and that the resulting empirical 2D surface maps can be directly compared to the atomic diffusion theory.

However, for a number of years, attempts to compare theoretically computed ApBp-star abundance maps with real observations appear to have been largely futile due to a fundamental failure of modern diffusion calculations to generate the observed diversity of the surface locations of chemical spots. As summarised by \citet{alecian:2015} for the equilibrium diffusion model, all elements that are expected to exhibit a strongly non-uniform distribution across the stellar surface (Alecian specifically mentions Mg, Al, Ti, V, Cr, Mn, Fe, Co, Ni) are expected to accumulate at different depths only in the region of nearly horizontal magnetic field lines, i.e. at the magnetic equator for stars with dipole-like global field geometries. As of yet, no equivalent summary of the time-dependent diffusion predictions has been published. However, according to \citet{alecian:2017} as well as Figs.~\ref{strat_fe} and \ref{strat_cr} of our paper, at least Cr and Fe are, again, predicted to accumulate exclusively at the magnetic equator.

As noted by many observational studies \citep[e.g.][]{kochukhov:2004e,kochukhov:2017a,nesvacil:2012,silvester:2014a,silvester:2015}, such equatorial overabundance rings are exceedingly rare in real ApBp stars. The only well-established examples of such surface structures are the O map of $\varepsilon$~UMa \citep*{rice:1997} as well as the C and O distributions of HR\,3831 \citep{kochukhov:2004e}. All other cases of various chemical elements mapped in about 40 ApBp stars show different distributions, often with spots located exactly at, or close to, the magnetic poles, where no element accumulation at any depth should occur according to either equilibrium or time-dependent version of the atomic diffusion theory.

Let us emphasise that this tension between theoretical predictions and observations is not engendered by the recent Doppler imaging results. Surface mapping based on the inverse problem solution is only one (through the most detailed and ambitious) among many techniques of obtaining spatial information concerning the chemical spot locations. Other, simpler and more direct, techniques have been extensively used in the past to infer spot properties for many ApBp stars. One can find numerous examples of historic Ap-star studies which discussed element spot locations relative to the magnetic field geometry with the help of comparison of the phase curves of radial velocity and equivalent widths of spectral lines, on the one hand, and the longitudinal magnetic field variation, on the other hand \citep[e.g.][]{stibbs:1950,pyper:1969,megessier:1975,mathys:1991,polosukhina:1999}. These straightforward analyses unambiguously demonstrate the ubiquitous presence of multiple groups of variable lines with different, sometimes anti-phase, rotational variations. A typical example of this behaviour is illustrated in Fig.~\ref{ewbz}, where we compare the longitudinal magnetic field measurements by \citet{silvester:2012} for the prototypical Ap star $\alpha^2$~CVn with the equivalent width variation of Cl and Cr lines employed for the DI mapping of this star by \citet{silvester:2014a}. These curves make it abundantly clear that the surface distributions of these two chemical elements cannot be the same. In this case, it is evident that the Cl distribution must be dominated by a single, large spot coinciding with the negative magnetic pole of the star, fully in line with the DI results published by \citet{silvester:2014a}. 

To conclude, our evaluation of the time-dependent atomic diffusion calculations by SA16 demonstrates two major weaknesses of their model. First, their calculations fail to produce mean abundance accumulation compatible with large Fe-peak element overabundances observed in late-B and early-A magnetic ApBp stars. Second, the prediction of similar horizontal abundance distributions, with accumulation at the magnetic equator for all elements, stands in fundamental conflict with the well-established diversity of the surface abundance structures in ApBp stars. It appears that, despite notable advances in the numerical treatment of polarised radiative transfer and its coupling with the atomic diffusion processes in a magnetised stellar atmosphere, the time-dependent diffusion model by SA16 does not reach the level of realism necessary for interpretation of observations of real stars. 

\section*{Acknowledgments} 
O.K. acknowledges financial support from the Knut and Alice Wallenberg Foundation, the Swedish Research Council, and the Swedish National Space Board. We thank the referee for detailed comments and many helpful suggestions which improved the clarity of our presentation.

\label{lastpage}

\end{document}